# Interpolated Topology Change in a Spin Cobordism and the Chiral Weyl Curvature Diagnostic

Keith Andrew[1], Eric Steinfelds[1,2], Kristopher Andrew[3]
Physics and Astronomy, Western Kentucky University
Bowling Green, KY 42101 USA
Department of Computational and Physical Sciences, Carroll University
Waukesha, WI 53186 USA
Science Department, Schlarman Academy
Danville, IL 61832 USA

**Abstract**

Topology change in Lorentzian quantum gravity demands geometric regulators that control curvature, spin structure, and chirality during nontrivial interpolations. We develop a framework for regulated topology change based on smooth Lorentzian spin cobordisms with interpolating metrics, allowing a transient failure of global hyperbolicity while preserving smoothness, Lorentz signature, and spin compatibility. Within this framework we introduce the Chiral Weyl Curvature Diagnostic, a curvature-based functional that weights topology-changing geometries by conformal curvature, spin admissibility, and topological complexity. The diagnostic functional is built from Weyl curvature invariants and includes a parity-odd dual Weyl term that is sensitive to geometric chirality. Spin consistency is enforced via a Stiefel–Whitney constraint, ensuring that only physically admissible cobordisms contribute. As an example, we construct a smooth spin cobordism interpolating between a Morris–Thorne wormhole and asymptotically flat Minkowski spacetime. In the throat region the curvature response is shown to be Weyl-dominated, and the parity-odd Weyl contribution sharply distinguishes chiral knotted embeddings while vanishing for amphichiral configurations. We then show that braids provide the natural language of throat dynamics: evolving wormhole throats trace time-dependent braid movies, with elementary braid generators representing the fundamental topological operations of the cobordism. Replacing crossing number by a braid-based complexity refines the diagnostic functional to operate at the level of these elementary exchanges and extends it naturally to multithroat and networked configurations.

## I. Introduction

Topology change in Lorentzian spacetimes has long been constrained by classical results due to Geroch, Tipler, and Sorkin, [1, 2, 3, 4] which show that smooth transitions between inequivalent spatial topologies generically require singularities, causality violations, or a breakdown of global hyperbolicity [5, 6, 7, 8]. As a consequence, much of the literature on spacetime topology change has either (i) relied on Euclidean methods and instanton constructions [9, 10], or (ii) restricted attention to purely topological or index-theoretic classifications that abstract away the Lorentzian causal and geometric structure relevant to semiclassical gravity [11, 12, 13, 14, 15].

In this work, we adopt a different perspective. Rather than proposing new dynamics or attempting to evade the classical obstructions by fiat, we focus on the geometric regulation and classification of topology-changing configurations within Lorentzian geometry itself. Specifically, we construct smooth, time-dependent Lorentzian metrics on spin cobordisms that interpolate between wormhole-bearing and topologically trivial spatial manifolds [16, 17, 18]. The interpolation preserves smoothness, Lorentz signature, and spin compatibility throughout, while allowing a localized and transient loss of global hyperbolicity in the interpolating region. This controlled relaxation is sufficient to evade the Geroch–Tipler obstruction without introducing curvature singularities or closed timelike curves.

The central objective of the paper is not to define a complete theory of topology change, but to develop covariant geometric diagnostics capable of distinguishing and weighting admissible topology-changing geometries in a semiclassical setting. An analysis of explicit interpolating spin cobordisms shows that curvature near wormhole throats is dominated by conformal (Weyl) contributions (see Sec. IV) [19, 20, 21]. Wormhole throats are used here as a concrete and geometrically controlled realization of localized topology change; the curvature diagnostic itself is not restricted to wormhole geometries.

Building on this analysis, we introduce the Chiral Weyl Curvature Diagnostic: a covariant functional constructed from the Weyl tensor and its parity-odd dual contraction. The curvature diagnostic operates entirely within Lorentzian geometry and has three defining features. First, it is dominated by conformal curvature in regions of nontrivial topology, providing sensitivity to localized geometric structure associated with wormhole throats. Second, through the inclusion of a parity-odd Weyl channel, it is sensitive to geometric chirality, changing sign under orientation reversal and vanishing for mirror-symmetric (amphichiral) embeddings. Third, spin admissibility is enforced via the vanishing of the second Stiefel–Whitney class, ensuring that only spin-compatible cobordisms[1] contribute [22, 23].

As a concrete realization of the framework, we construct smooth Lorentzian spin cobordisms [24, 25] interpolating between Morris–Thorne wormholes and asymptotically flat spacetimes [26, 27]. We demonstrate that the parity-odd Weyl contribution sharply distinguishes chiral knotted throat embeddings while vanishing for amphichiral configurations such as the figure-eight knot [28, 29]. We further argue that evolving wormhole throats are naturally described by braid movies: time-dependent braid representations in which elementary braid generators correspond to local topological exchanges during the cobordism [30, 31, 32]. Replacing knot crossing number with a braid-generator–based complexity refines the normalization of the curvature diagnostic without altering its curvature structure and extends its applicability to time-dependent, multithroat configurations.

The result is a curvature-level, chirality-sensitive, spin-filtered measure of the geometric cost of Lorentzian topology change. It introduces no new equations of motion and makes no claims of universality. Instead, it provides a principled geometric weighting mechanism that complements existing Euclidean, index-theoretic, and anomaly-based approaches by operating directly within Lorentzian geometry and resolving chirality at the level of curvature. Throughout this work, the results are intended as geometric diagnostics and regulatory criteria for admissible Lorentzian topology change, rather than as new invariants, index theorems, or modifications of gravitational dynamics. This paper introduces a Lorentzian interpolating spin-cobordism framework for topology change and proposes a chirality-sensitive Weyl curvature functional as a diagnostic and semiclassical weighting of admissible geometries. By combining parity-odd Weyl contractions with spin admissibility and a controlled complexity normalization, the construction distinguishes embeddings related by mirror reversal while suppressing trivial or collapsed topologies. Rather than defining a topological invariant or a full quantum measure, the curvature diagnostic provides a curvature-level selection principle that connects topology change, spin structure, and chirality within a purely geometric—and explicitly Lorentzian—setting. For clarity regarding scope, intent, and limitations, we summarize the role of the present work in Box 1 before introducing the technical framework.

**Box 1. Scope and Intended Role of the Present Work**

This paper develops a geometric framework for analyzing topology change in four-dimensional Lorentzian spacetimes. For clarity, we summarize below the scope and intended role of the results.

**What is established**

• **Lorentzian spin cobordisms with smooth interpolating metrics.**
We construct smooth Lorentzian metrics on spin cobordisms that interpolate between wormhole-bearing and topologically trivial spatial manifolds. Smoothness, Lorentz signature, and spin compatibility are preserved throughout, while global hyperbolicity is allowed to fail locally and transiently so that here no closed timelike curves arise because the interpolating region lacks global Cauchy surfaces without admitting causal loops.

• **Dominance of conformal curvature near wormhole throats.**
An analysis of curvature invariants shows that near wormhole cores the Weyl (conformal) curvature dominates over Ricci and scalar curvature contributions, motivating the use of Weyl-based geometric diagnostics.

---

1. A Lorentzian spin structure on the cobordism $\mathcal{M}$ exists if and only if $w_2(\mathcal{M}) = 0$. When this condition fails, no globally defined spin bundle or Dirac operator exists on $\mathcal{M}$. Since the present framework is intended to weight only spin-admissible geometries, non-spin cobordisms are suppressed identically by the selection factor $\chi_{spin}(\mathcal{M})$.

- **A chirality-sensitive curvature functional.**
We introduce a covariant functional constructed from the Weyl tensor and its parity-odd dual contraction. This functional distinguishes orientation-reversed geometric embeddings and vanishes for mirror-symmetric (amphichiral) configurations.

- **Geometric enforcement of spin admissibility.**
Spin compatibility is imposed through the vanishing of the second Stiefel–Whitney class of the cobordism, ensuring that only spin-admissible geometries contribute.

- **A diagnostic weighting for topology-changing configurations.**
The resulting functional provides a curvature-based diagnostic suitable for weighting or suppressing topology-changing geometries in semiclassical analyses, without introducing new dynamics.

- **Extension to time-dependent multithroat configurations.**
Evolving wormhole throats are naturally described by braid representations, and a braid-generator–based complexity provides a refinement of crossing-based normalization appropriate for time-dependent networks.

What is not claimed:

- No modification of Einstein dynamics or gravitational field equations is proposed.
- The curvature functional is not a topological invariant and does not compute an index or quantized charge.
- No claim is made regarding the necessity, prevalence, or dominance of topology change in quantum gravity.
- Energy-condition violations may occur during the interpolating phase; only smoothness and spin compatibility are enforced.
- The work does not provide a nonperturbative definition of the gravitational path integral.
- Fermionic spectral flow, η-invariants, and anomaly calculations are not developed here.

## II. Lorentzian Spin Cobordisms and Admissibility Conditions

We briefly summarize the geometric framework required for Lorentzian topology change and fix notation. Detailed reviews of cobordism theory may be found elsewhere and will not be repeated here [33]. Two oriented $n$-dimensional manifolds $M_1$ and $M_2$ are cobordant if there exists a compact $(n+1)$-dimensional manifold $W$ such that
$$\partial W = M_1 \sqcup (-M_2). \tag{1}$$

When $W$ admits a Lorentzian metric whose restriction to each boundary is Lorentzian, we refer to $W$ as a Lorentzian cobordism between $M_1$ and $M_2$. In the present work, $M_1$ represents a compact spatial slice containing one or more wormhole throats, while $M_2$ is a compactified trivial topology (e.g. $S^3$). The cobordism $W$ is equipped with a smooth Lorentzian metric that interpolates between these boundary geometries. We do not require global hyperbolicity of $W$; instead, we allow a localized region in which no global Cauchy surface exists, in accordance with the Geroch–Tipler obstruction. Outside this region, the spacetime remains globally hyperbolic. Because fermionic fields play a central role in quantum gravity, we restrict attention to spin cobordisms. A spin structure exists on $W$ if and only if the second Stiefel–Whitney class vanishes,
$$w_2(W) = 0. \tag{2}$$

Throughout, the analysis is restricted to Lorentzian spin-admissible cobordisms; topology-changing interpolations excluded by the standard Lorentzian spin selection rules are assumed absent from the outset [34]. This restriction ensures that Dirac spinors may be consistently defined on the cobordism and its boundaries. In the examples considered, both the initial wormhole geometry and the final trivial geometry admit spin structures, and the interpolating cobordism is taken to be spin-compatible. The admissibility criteria imposed in this paper are therefore purely geometric:
1. The metric on $W$ is smooth and Lorentzian.
2. Lorentz signature is preserved throughout the interpolation.
3. No closed timelike curves are introduced.

4. Global hyperbolicity may fail only transiently in the interpolating region.
5. The spin condition $w_2(W) = 0$ is satisfied everywhere.

From a global perspective, the existence of a Lorentzian interpolation between prescribed initial and final spatial hypersurfaces is known to be governed by cobordism-theoretic conditions rather than by local curvature alone. In particular, classical results on Lorentzian cobordisms show that the existence of a Lorentzian metric with spacelike boundary is equivalent to the existence of a nowhere-vanishing line field transverse to the boundary components, with additional parity constraints arising when a Spin structure is required. These results provide necessary and sufficient *admissibility conditions* for topology-changing histories, expressed in terms of global invariants such as Euler (semi-)characteristics or related kink-type indices. However, such theorems are inherently global and topological in nature: they determine which interpolations are allowed but are silent on the local geometric structure of the interpolation region itself. Moreover, compact Lorentzian cobordisms satisfying these hypotheses are known to exhibit generic causal pathologies, motivating the use of weak or localized[2] Lorentzian interpolations in which global hyperbolicity may fail transiently without introducing closed timelike curves.

In this setting, the Weyl Curvature Chiral Diagnostic introduced here should be understood as a complementary, local refinement: once admissibility is satisfied at the cobordism level, the parity-odd Weyl invariant $C_{abcd}{}^*C^{abcd}$ provides a covariant measure of the chiral and conformal curvature content generated during the interpolation. The diagnostic thus does not replace cobordism-based existence criteria but rather supplies a geometric weighting that distinguishes dynamically chiral topology-changing histories from those that are topologically allowed but curvature-neutral. Within this class of admissible Lorentzian spin cobordisms, we seek geometric diagnostics that distinguish and weight topology-changing configurations. In the sections that follow, we construct interpolating metrics satisfying these conditions, analyze their curvature structure, and introduce a Weyl-based functional that captures both conformal curvature concentration and geometric chirality.

### III. Interpolating Lorentzian Spin Geometry

We construct a family of smooth Lorentzian metrics that interpolate between a wormhole-bearing spatial topology and a topologically trivial configuration, defined on a four-dimensional spin cobordism $W$. The interpolation parameter $u \in [0,1]$ labels the progression from the initial wormhole geometry ($u = 0$) to the trivial geometry ($u = 1$).

### A. Metric Ansatz

We adopt a time-dependent metric of the form
$$ds^2 = -dt^2 + a^2(t)\, g_{ij}(x;u)\, dx^i dx^j, \tag{3}$$

where $a(t)$ is a smooth cosmological scale factor and $g_{ij}(x;u)$ is a family of spatial metrics interpolating between a multithroat wormhole geometry and a compactified trivial topology. In the absence of wormholes, this reduces to the standard Friedmann–Lemaître–Robertson–Walker metric, while for static $a(t)$ it reproduces a multithroat generalization of the Morris–Thorne wormhole. For a single throat, the spatial metric may be written in the form
$$g_{ij}(x;u)\, dx^i dx^j = \frac{dr^2}{1-b(r,u)/r} + r^2(d\theta^2 + \sin^2\theta\, d\phi^2), \tag{4}$$

with an interpolation of the shape function
$$b(r,u) = (1-u)\, b_{\mathrm{WH}}(r), \tag{5}$$

so that the throat closes smoothly as $u \to 1$. Generalization to multiple interacting throats is achieved by introducing localized envelope functions and interaction terms; these details do not affect the arguments below and will be suppressed. The interpolation is constructed so that all metric components and their derivatives remain smooth functions of $(x, u)$.

### B. Lorentz Signature and Causality

---

2. Lorentzian topology change has also been discussed in settings involving non-Hausdorff or causally branching structures; because the Weyl Curvature Chiral Diagnostic is defined locally on smooth Lorentzian regions and does not depend on global manifold properties, these subtleties play no role in the present analysis.

The metric retains Lorentzian signature $(-,+,+,+)$ for all $u \in [0,1]$. No metric component changes sign, and no curvature scalar diverges at finite $u$. The interpolating metrics considered here are constructed so as not to introduce closed timelike curves. Global hyperbolicity holds in the initial and final geometries but is not enforced in the interpolating region. The absence of a global Cauchy surface during the transition is essential to evade the Geroch–Tipler obstruction and is localized in $u$. Outside this region spacetime is globally hyperbolic.

### C. Spin Structure

Both boundary geometries admit spin structures, and the interpolating manifold $W$ is chosen such that $w_2(W) = 0$. The spin structure extends smoothly across the cobordism, ensuring that fermionic fields are well-defined throughout the topology-changing process. No discontinuities or obstructions arise from the interpolation itself. This spin admissibility condition will later enter explicitly as a selection rule in the Weyl curvature diagnostic functional.

### D. Summary of Geometric Properties

The interpolating geometry constructed above satisfies the following properties:

- A smooth Lorentzian metric on the spin cobordism $W$.
- Fixed Lorentz signature throughout the interpolation.
- Absence of closed timelike curves.
- Temporary loss of global hyperbolicity localized to the interpolating region.
- A globally well-defined spin structure.

These properties ensure that the geometry is admissible within semiclassical gravity while allowing nontrivial topology change. All curvature invariants remain finite throughout the interpolation. Fig. (1) illustrates the evolution of representative components of the Weyl curvature diagnostic as functions of the interpolation parameter $u$. A comparative analysis of curvature components and their relative dominance is given in Sec. IV.

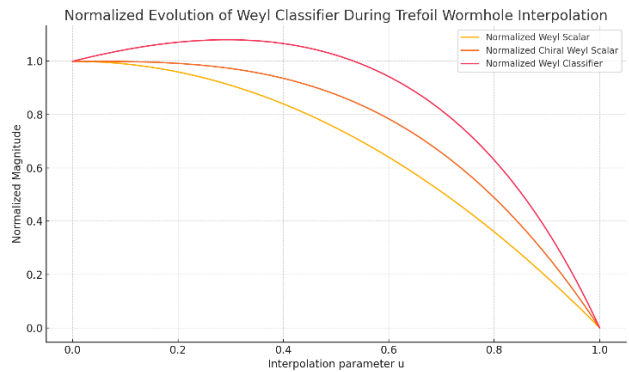

**Fig. 1.** Normalized evolution of the Weyl curvature diagnostic during trefoil wormhole interpolation. Shown are the scalar Weyl contribution (yellow), the parity-odd Weyl contribution (orange), and the combined diagnostic (red) as functions of the interpolation parameter $u$, ranging from the initial knotted wormhole geometry ($u=0$) to flat spacetime ($u=1$).

In the next section, we analyze the curvature response of this interpolating family to identify which geometric invariants are most sensitive to localized topology change.

### IV. Curvature Response and Weyl Dominance

To motivate a curvature-based diagnostic for topology change, we analyze the behavior of standard curvature scalars along the interpolating Lorentzian spin cobordism constructed in Section III. Our goal is to identify which curvature components dominate near wormhole throats and are therefore most sensitive to localized, topologically nontrivial geometry. Throughout this section we treat multithroat interactions and knot deformations as localized perturbations about a smooth background geometry and build to leading nontrivial order.

### A. Curvature Scalars for the Interpolating Geometry

We consider the Ricci scalar $R$, the Kretschmann scalar

$$K = R_{\mu\nu\rho\sigma}R^{\mu\nu\rho\sigma}, \tag{6}$$

and the squared Weyl invariant

$$C^2 = C_{\mu\nu\rho\sigma}C^{\mu\nu\rho\sigma}, \tag{7}$$

evaluated on the interpolating metric family. For the class of metrics considered here, the Ricci scalar decomposes schematically as

$$R = R_{\text{bg}} + R_{\text{loc}} + R_{\text{int}}, \tag{8}$$

where $R_{\text{bg}}$ arises from cosmological expansion, $R_{\text{loc}}$ encodes the localized wormhole curvature, and $R_{\text{int}}$ contains interaction and deformation terms associated with multithroat structure. Near the wormhole throat, $R_{\text{bg}}$ is subleading, while $R_{\text{loc}}$ remains finite and smooth throughout the interpolation. The Kretschmann scalar exhibits a similar decomposition, with the dominant localized contribution scaling quadratically with tidal curvature near the throat. While $K$ provides a measure of overall curvature intensity, it does not distinguish between conformal and Ricci curvature contributions.

### B. Weyl Curvature Localization

The Weyl tensor isolates the trace-free, conformal part of the curvature and is therefore insensitive to local matter content and cosmological background. For the interpolating wormhole geometry, the squared Weyl invariant takes the schematic form

$$C^2 \sim \sum_n f_n(r) C_n^2 + \sum_{m \neq n} C_{mn}^2, \tag{9}$$

where the envelope functions $f_n(r)$ localize curvature near each throat and the interaction terms encode knotting, linking, or twisting between throats. Near the wormhole cores, the Weyl contribution dominates over both Ricci and mixed curvature terms. Explicitly, the ratios

$$\frac{R}{C^2}, \frac{K}{C^2} \tag{10}$$

are suppressed near the throat and vanish in the limit of strong localization. This behavior reflects the fact that wormhole throats are regions of high tidal deformation but comparatively low scalar curvature. As a result, conformal curvature provides the most sensitive probe of localized geometric structure associated with nontrivial topology. Appendix C shows the curvature scalars and energy conditions for this cobordism.

### C. Sensitivity to Topological Embedding

Crucially, the Weyl tensor is sensitive not only to the presence of a throat but also to its *embedding*. Knotting or twisting of a wormhole throat introduces anisotropic tidal deformations that are captured by the Weyl curvature but largely invisible to scalar Ricci diagnostics. This sensitivity is reflected in evaluations of $C^2$ for representative knotted embeddings. For fixed throat scale, more compact or tightly embedded knots yield larger Weyl curvature densities, while extended or amphichiral embeddings produce reduced conformal response. The Weyl invariant thus naturally encodes both geometric concentration and embedding complexity. However, the scalar Weyl invariant remains parity even. Mirror-related embeddings yield identical contributions to $C^2$, indicating that scalar conformal curvature alone cannot distinguish geometric chirality. While the examples employ simple matter regulators (e.g. a cosmological constant or localized envelope functions), the dominance of conformal (Weyl) curvature near wormhole throats is a geometric feature of the interpolating Lorentzian cobordism and does not rely on the detailed form of the supporting stress–energy tensor.

**D. Implications for a Curvature-Based Diagnostic**

The analysis above establishes three key points:

1. Weyl curvature dominates near wormhole throats in the interpolating geometry.
2. Conformal curvature is maximally sensitive to localized, topologically nontrivial structure.
3. Scalar curvature invariants are blind to chirality, motivating an extension beyond parity-even diagnostics.

These observations motivate a curvature-based Weyl curvature diagnostic built from the Weyl tensor rather than Ricci curvature or global topological invariants. To distinguish mirror-related embeddings and detect geometric handedness, such a curvature diagnostic must incorporate a parity-odd curvature channel. The overall Weyl dominance is exhibited in Fig. (2) and shows that as u increases, curvature invariants decay toward zero, reflecting a smooth topological transition from a high-curvature wormhole core to flat spacetime.

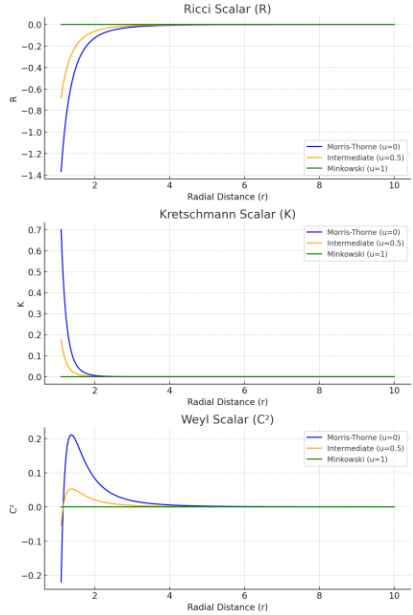

**Fig. 2.** Weyl Dominance in radial profiles of curvature scalars for three stages of the interpolating cobordism: initial wormhole geometry ($u = 0$), intermediate geometry ($u = 0.5$), and asymptotic Minkowski limit ($u = 1$). Shown are the Ricci scalar $R$, the Kretschmann scalar $K$, and the squared Weyl invariant $C^2$.

Fig. (2) shows that, across the interpolating family, the conformal (Weyl) curvature remains the dominant localized contribution near the throat, while Ricci and scalar curvature terms are comparatively suppressed and decay more rapidly as the geometry relaxes toward the trivial end state. Although the Ricci scalar attains larger absolute values near the throat due to the matter content supporting the wormhole, we see that the conformal (Weyl) curvature is the component that localizes most sharply to the throat and decays independently of the background, identifying it as the geometric carrier of topology-sensitive structure. As such Ricci can be large without being topologically diagnostic. Among all local curvature diagnostics available in Lorentzian geometry, the parity-odd Weyl contraction is essentially unique among local, polynomial curvature scalars in four-dimensional Lorentzian geometry in being simultaneously conformally invariant, orientation-sensitive, spin-compatible, and independent of matter content, making it a natural carrier of chiral information during topology change.

**E. Structural Uniqueness of the Parity-Odd Weyl Channel**

The curvature analysis of Sec. IV establishes that conformal curvature dominates the localized geometric response near wormhole throats in smooth Lorentzian spin cobordisms. This observation motivates the use of Weyl curvature as a diagnostic of topology change. In this subsection, we argue that the parity-odd Weyl contraction introduced below is not merely a convenient extension of this diagnostic but is structurally privileged among local Lorentzian curvature scalars.

In four dimensions, local scalar densities constructed from the Riemann tensor fall into a small number of classes. Scalars built from the Ricci tensor or Ricci scalar necessarily couple to matter content and volume change, and therefore conflate localized geometric structure with stress–energy contributions. Quadratic Riemann invariants such as the Kretschmann scalar mix conformal and Ricci curvature and are insensitive to orientation. Topological densities, such as the Euler and Pontryagin classes, are total derivatives and encode only global information, rendering them unsuitable as local geometric regulators along an interpolating cobordism.

By contrast, the Weyl tensor isolates the trace-free, conformally invariant part of the curvature and is maximally sensitive to anisotropic tidal deformation. For localized throat geometries, this makes Weyl curvature the natural carrier of shape information associated with nontrivial topology, as demonstrated in Sec. IV.B. However, scalar Weyl invariants remain parity even and therefore cannot distinguish mirror-related embeddings.

The contraction of the Weyl tensor with its Hodge dual provides a distinguished parity-odd scalar density. Among local curvature scalars in Lorentzian geometry, this quantity is essentially unique in satisfying the following properties simultaneously:

(i) **Conformal invariance:** it depends only on the conformal geometry and is insensitive to overall volume rescalings;

(ii) **Orientation sensitivity:** it changes sign under orientation reversal and therefore detects geometric chirality;

(iii) **Local covariance:** it is a scalar constructed pointwise from the curvature without reference to global topology or boundary data;

(iv) **Matter independence:** it is blind to Ricci curvature and therefore decoupled from the detailed form of the stress–energy tensor;

(v) **Spin compatibility:** it is well defined on spin-admissible Lorentzian cobordisms and naturally suppressed when spin selection rules exclude the geometry.

These constraints sharply restrict the available curvature diagnostics. In particular, any local scalar sensitive to chirality must involve the Levi–Civita tensor and therefore be parity odd, while any scalar insensitive to matter content must be purely Weyl. The dual Weyl contraction is thus singled out as the unique local curvature channel capable of encoding geometric handedness during Lorentzian topology change.

From this perspective, the appearance of a parity-odd Weyl contribution is not an arbitrary refinement but a structural consequence of the geometric requirements imposed by Lorentzian topology change. The resulting scalar does not define a topological invariant and is not quantized; rather, it measures the localized concentration of chiral conformal curvature along the cobordism. As the topology trivializes and the geometry relaxes toward conformal flatness, this channel must decay, consistent with the behavior exhibited in Secs. IV and V.

This structural uniqueness motivates the introduction of the chiral Weyl curvature functional in the next section as a diagnostic and weighting mechanism for admissible topology-changing geometries, rather than as an additional assumption or model-dependent choice.

### IV.E. Curvature-Channel Selection Rule for Chiral Topology Change

The analysis above shows that parity-odd conformal curvature responds selectively to certain geometric features of topology-changing Lorentzian cobordisms. This motivates the introduction of a selection rule associated not with the existence of topology change itself, but with the activation of the parity-odd Weyl curvature channel used in the Weyl curvature diagnostic.

Importantly, the selection rule formulated here is diagnostic rather than prohibitive. It does not assert which topology changes are dynamically realized or globally admissible. Instead, it identifies the geometric conditions under which a topology-changing history can contribute nontrivially to the parity-odd Weyl diagnostic.

**Selection Rule (Parity-Odd Weyl Admissibility).**

*Let $W$ be a smooth Lorentzian cobordism interpolating between spatial slices $\Sigma_{\text{in}}$ and $\Sigma_{\text{out}}$. A necessary condition for $W$ to contribute nonvanishing parity-odd Weyl curvature is that the topology-changing history be both spin-admissible and chirally unpaired. In particular:*

1. *$W$ must admit a spin structure compatible with the interpolating geometry;*
2. *the oriented local topology-changing events occurring along $W$ must not decompose into orientation-reversed pairs.*

*If either condition fails, the spacetime-integrated parity-odd Weyl density vanishes identically.*

This rule follows directly from the structure of the parity-odd Weyl contraction. Spin inadmissibility obstructs the definition of orientation-sensitive geometric data, while orientation-reversed pairing of local events leads to exact cancellation in the parity-odd channel. In this sense, parity-odd Weyl curvature functions as a filter that is blind to topology change, lacking intrinsic geometric handedness.

Several clarifications are in order. First, the selection rule does not restrict which Lorentzian cobordisms may exist, nor does it impose global consistency conditions such as causal regularity or global hyperbolicity. A topology-changing geometry that violates the conditions above may still be realized as a Lorentzian spacetime; it simply leaves no imprint in the parity-odd Weyl diagnostic.

Second, the selection rule is channel-specific. It applies only to the parity-odd conformal curvature channel considered here. Other curvature diagnostics—parity-even scalars, Ricci-based measures, or spectral quantities—may respond differently to the same geometry. The rule therefore does not constitute a general prohibition on topology change, but rather a statement of geometric sensitivity.

Finally, the selection rule is local and structural rather than dynamical. It does not rely on equations of motion, energy conditions, or quantization, and it is independent of matter content. Its role is to clarify which topology-changing histories can be detected by the Weyl-chirality curvature diagnostic, and why amphichiral or spin-obstructed histories are necessarily invisible to this channel.

This perspective places the Weyl curvature diagnostic in a complementary position to global consistency theorems and anomaly-based approaches. Where the latter address whether topology change can occur or how quantum fields respond to it, the present selection rule identifies which topology-changing geometries possess the intrinsic conformal chirality required to register in a parity-odd curvature diagnostic [35][3].

---

3. This selection rule should not be confused with global obstruction theorems for Lorentzian topology change, such as those due to Geroch, Tipler, or Hawking, nor with analyses of causal or foliation pathologies associated with kinked time functions (e.g. Chamblin [33]). Those results address the existence, global consistency, or causal

In the next section, we introduce a chiral extension of the Weyl curvature diagnostic, based on the contraction of the Weyl tensor with its dual. This construction yields a covariant, chirality-sensitive functional that operates entirely within Lorentzian geometry and naturally filters topology-changing configurations by conformal curvature, embedding complexity, and spin admissibility. In summary, smooth Lorentzian wormhole cobordisms exhibit Weyl-dominated curvature near the throat, while chirality enters only through parity-odd Weyl contractions, providing the geometric basis for the Weyl functional introduced in Sec. V. For illustrative purposes the examples in Appendix C employ a constant-curvature background to cleanly separate localized throat dynamics from infrared geometry.

## V. Chiral Weyl Curvature Diagnostic

With these results from Sec. IV, we now define such a functional. Here we use the term curvature diagnostic in a diagnostic sense: a covariant geometric functional that discriminates admissible configurations, rather than a complete invariant of topology or homotopy class. Before defining the Weyl curvature diagnostic, we clarify its relation to anomaly-based and spectral approaches.

### A. Relation to Anomalies and Spectral Asymmetry

Parity-odd structures are most commonly encountered in gravitational physics through fermionic anomalies, η-invariants, and index-theoretic constructions. These approaches play a central role in understanding chiral symmetry breaking, spectral flow, and topological charge in quantum field theory on curved spacetimes. It is therefore natural to ask how the parity-odd Weyl curvature channel introduced here relates to these well-established frameworks.

The distinction is both conceptual and structural. Anomalies and η-invariants are intrinsically quantum objects: they depend on the quantization of fermionic fields, on the choice of operator and boundary conditions, and on the global spectral properties of the Dirac operator. Their evaluation typically requires a complete account of the fermion content of the theory and yields global or semi-global quantities tied to spectral asymmetry. As such, anomaly-based diagnostics do not exist independently of quantum matter fields and do not admit a purely geometric definition at the classical level.

By contrast, the parity-odd Weyl curvature functional introduced in this work is defined entirely within classical Lorentzian geometry. It exists prior to quantization, requires no reference to fermionic degrees of freedom, and depends only on local curvature structure and orientation. Its definition makes no appeal to eigenvalue spectra, boundary conditions, or regularization schemes. The functional therefore characterizes geometric chirality directly at the level of the spacetime manifold and its spin-admissible cobordism, rather than through the response of quantum fields propagating on that background.

From this perspective, the two approaches are complementary rather than competing. The parity-odd Weyl channel measures the presence and spacetime support of chiral conformal curvature during topology change, while anomaly and spectral-flow calculations quantify how quantum fermions respond to such chiral geometric environments. In particular, while anomaly inflow or η-invariants capture integrated spectral asymmetry after quantization, the Weyl curvature functional identifies where and when chiral geometric structure is available to source such effects.

In this sense, the parity-odd Weyl channel may be viewed as the geometric precursor of chiral spectral asymmetry, rather than its quantum manifestation. It supplies a curvature-level diagnostic that exists independently of matter content and persists even in the absence of fermions, while naturally providing the geometric substrate upon which anomaly-based and spectral approaches may later act.

The present work therefore does not replace anomaly or index-theoretic analyses, nor does it attempt to derive them. Instead, it isolates a purely Lorentzian, curvature-based notion of chirality that complements those frameworks by

---

structure of topology-changing spacetimes. The present rule is strictly channel-specific: it assumes a smooth Lorentzian cobordism and characterizes only whether such a geometry can register in the parity-odd Weyl curvature diagnostic.

operating at an earlier, pre-quantum stage. A detailed investigation of how the parity-odd Weyl channel couples to fermionic spectral flow on spin cobordisms lies beyond the scope of this paper and will be addressed elsewhere.

**Definition 1 (Chiral Weyl Curvature Diagnostic : Definition and Properties).**
Let $W$ be a smooth four-dimensional Lorentzian spin cobordism interpolating between spatial manifolds $\Sigma_1$ and $\Sigma_2$, satisfying the admissibility conditions of Sec. II. Define the scalar Weyl functional

$$\mathcal{W}[g] \;=\; \int_W C_{\mu\nu\rho\sigma}\, C^{\mu\nu\rho\sigma}\, \sqrt{-g}\, d^4x, \tag{11}$$

and the parity-odd Weyl functional

$$\mathcal{W}_\chi[g] \;=\; \int_W C_{\mu\nu\rho\sigma}{}^*\, C^{\mu\nu\rho\sigma}\, \sqrt{-g}\, d^4x, \tag{12}$$

where $*C^{\mu\nu\rho\sigma}$ is the Hodge dual of the Weyl tensor.

Then the normalized functional

$$\boxed{\mathcal{C}_W[g] \;=\; \tfrac{1}{\mathcal{N}}\bigl(\mathcal{W}[g] \;+\; \lambda\, \mathcal{W}_\chi[g]\bigr)\, \chi_{spin}\bigl(w_2(W)\bigr)} \tag{13}$$

defines a covariant curvature-based functional for Lorentzian topology change, where:

- $\lambda$ controls the relative weight of the parity-odd channel,

- $\lambda$ is a dimensionless weighting factor controlling the relative contribution of the parity-odd Weyl channel; in practice one expects $\lambda = \mathcal{O}(1)$ in Planck units, with $\lambda = 0$ suppressing chirality sensitivity and $\lambda \sim 1$ yielding comparable weighting of parity-even and parity-odd curvature contributions,

- $\mathcal{N}$ is a geometric normalization depending on throat arclength and topological complexity,

- $\chi_{spin}(w_2)$ is defined to vanish unless the second Stiefel–Whitney class of $W$ vanishes, where we use $\chi_{spin}(w_2)$ as a shorthand for a binary admissibility factor enforcing spin compatibility, not as a continuous function of a cohomology class.

The normalization factor $\mathcal{N}$ is chosen to depend only on smooth geometric quantities—such as proper throat arclength and a bounded measure of topological complexity—and is assumed to remain finite throughout the cobordism, so that vanishing or enhancement of $W$ reflects curvature structure rather than normalization artifacts. Note that while $\lambda$ is treated here as a phenomenological weighting parameter, it is suggestive to interpret it as encoding the relative efficiency with which parity-odd conformal curvature sources fermionic spectral asymmetry on the spin cobordism, a connection we defer to future work.

**Properties:**

The Weyl functional $\mathcal{C}_W$ satisfies:

1. Covariance
   It is invariant under diffeomorphisms and independent of coordinate choices.

2. Weyl Dominance
   Near wormhole throats, $\mathcal{C}_W$ is dominated by conformal curvature contributions, consistent with Section IV.

3. Spin Admissibility
   Contributions from non-spin cobordisms are identically suppressed.

4. Chirality Sensitivity
   The parity-odd term changes sign under orientation reversal and vanishes for mirror-symmetric embeddings.

For example, cobordisms involving an odd self-intersection of throat world-sheets fail the spin condition and are automatically suppressed by the Weyl functional. In Fig. (3) we plot the functional for several knot types and see that the scalar captures the local chiral curvature induced by the topological handedness of each knot embedded in the wormhole throat

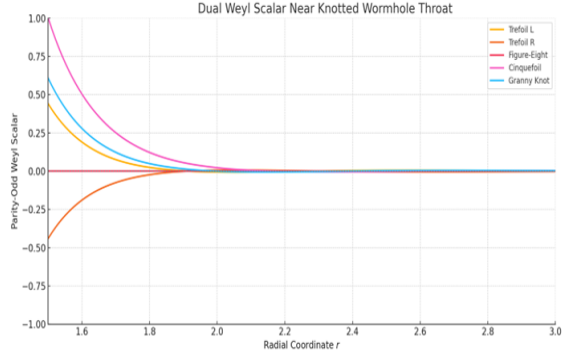

**Fig. 3.** Weyl and parity-odd Weyl curvature near knotted wormhole throats. The dual Weyl scalar is shown as a function of radial distance for left- and right-handed trefoil knots, the figure-eight knot, the cinquefoil knot, and the granny knot. Mirror-related embeddings exhibit opposite-sign responses, while the amphichiral figure-eight configuration yields negligible parity-odd curvature.

The following proofs are included for completeness and to fix conventions; the results themselves follow directly from standard properties of the Weyl tensor, orientation reversal, and spin admissibility, and are geometrically immediate.

**Lemma 1 (Parity-odd uniqueness).**
*Among local, covariant curvature scalars constructed from the Riemann tensor and its dual, the pseudoscalar*

$$\mathcal{W}_* = \int C_{abcd} \, \tilde{C}^{abcd} \, d^4x$$

*is the unique term that (i) vanishes on parity-symmetric embeddings, (ii) distinguishes mirror-related Lorentzian cobordisms, and (iii) remains insensitive to Ricci-sector redefinitions under smooth interpolations.*

*Proof.* Scalar contractions of the Weyl tensor are parity-even unless one factor is dualized. Ricci-dependent pseudoscalars reduce to total derivatives or vanish under spin-admissible cobordisms.

**Corollary 1 (Chirality Detection)**

*For a wormhole throat embedded with a chiral knot K,*

$$\mathcal{W}_*[K] \neq 0, \tag{14}$$

*and*

$$\mathcal{W}_\star[K] = -\mathcal{W}_\star[\bar{K}], \tag{15}$$

where $\bar{K}$ denotes the mirror embedding. For amphichiral embeddings,

$$\mathcal{W}_\star = 0. \tag{16}$$

*Proof (sketch).*
Let $\mathcal{M}$ be an oriented Lorentzian spin cobordism and let $\widetilde{\mathcal{M}}$ denote the same manifold equipped with the opposite orientation, corresponding to the mirror embedding $\bar{K}$. The scalar Weyl density $C_{abcd}C^{abcd}$ depends only on metric contractions and is therefore invariant under orientation reversal. By contrast, the dual Weyl tensor $*C_{ab}{}^{cd} = \frac{1}{2}\epsilon_{ab}{}^{mn}C_{mn}{}^{cd}$ involves the Levi–Civita tensor $\epsilon_{abcd}$, which changes sign under orientation reversal. As a result, the parity-odd density $C_{abcd}{}^*C^{abcd}$ reverses sign under $K \mapsto \bar{K}$, implying

$$\mathcal{W}_\star[K] = -\mathcal{W}_\star[\bar{K}]. \tag{17}$$

For amphichiral embeddings, an orientation-reversing isometry identifies $K$ with $\bar{K}$, so the same configuration must satisfy $\mathcal{W}_\star = -\mathcal{W}_\star$, hence $\mathcal{W}_\star = 0$. The stated result for the normalized functional $\mathcal{W}$ follows immediately for $\gamma \neq 0$. □

**Corollary 2 (Suppression of Trivial Topology)**

*For interpolations terminating in a topologically trivial geometry with vanishing Weyl curvature,*

$$\mathcal{W} \to 0. \tag{18}$$

*Proof (sketch).*
On the trivial end state, the spacetime geometry is conformally flat by assumption, so the Weyl tensor vanishes identically: $C_{abcd} = 0$. Both the parity-even and parity-odd Weyl densities therefore vanish pointwise,

$$C_{abcd}C^{abcd} = 0, C_{abcd}{}^*C^{abcd} = 0. \tag{19}$$

Consequently, $\mathcal{W}_2 = \mathcal{W}_\star = 0$, and the normalized functional $\mathcal{W}$ vanishes as well, provided the geometric normalization factor remains finite. □

**Corollary 3 (Non-Monotonic Evolution)**

*Along smooth interpolating families $g(u)$, the normalized functional $\mathcal{C}_W[g(u)]$ need not vary monotonically with the interpolation parameter $u$. Transient enhancement may occur when geometric compression of throat arclength outpaces curvature decay.*

**Remark 1 (Not a Topological Invariant)**

The Weyl functional $\mathcal{C}_W$ is not a topological invariant. It depends on the detailed geometry of the cobordism and is sensitive to embedding, curvature localization, and scale. Its role is diagnostic and regulatory rather than classificatory in the homotopy-theoretic sense. While the functional resembles the Pontryagin density, it differs crucially by being purely Weyl and not a total derivative. Consequently, the functional is not quantized and does not compute an index but rather measures local chiral curvature concentration along the cobordism.

**Remark 2 (Normalization and Complexity)**

In practice, the normalization factor $\mathcal{N}$ may be expressed as

$$\mathcal{N} \sim L_{\text{throat}}\, \kappa, \tag{20}$$

where $L_{\text{throat}}$ is the proper arclength of the throat and $\kappa$ encodes topological complexity.

While knot crossing number provides a simple choice for $\kappa$, a braid-generator–based complexity yields a refinement better adapted to time-dependent multithroat configurations. This substitution alters no curvature terms and preserves the structure of the Weyl curvature diagnostic. Appendix B provides a model for braided throat normalization.

**Remark 3 (Role in Semiclassical Gravity)**

The Weyl functional $\mathcal{C}_W$ is naturally interpreted as a weighting functional for topology-changing configurations in semiclassical gravitational path integrals. It supplies a covariant geometric suppression mechanism based on conformal curvature, chirality, and spin compatibility, without introducing new dynamical equations.

The Chiral Weyl Wormhole functional provides a curvature-level, chirality-sensitive, and spin-filtered measure of the geometric cost of Lorentzian topology change. It operates entirely within Lorentzian geometry, distinguishes mirror-related embeddings, and vanishes smoothly as topology trivializes. In doing so, it supplies a principled geometric regulator for topology change compatible with semiclassical gravity. Fig.(4) shows the functional magnitude across the cobordism and Appendix A gives the cohomology flow of the cobordism.

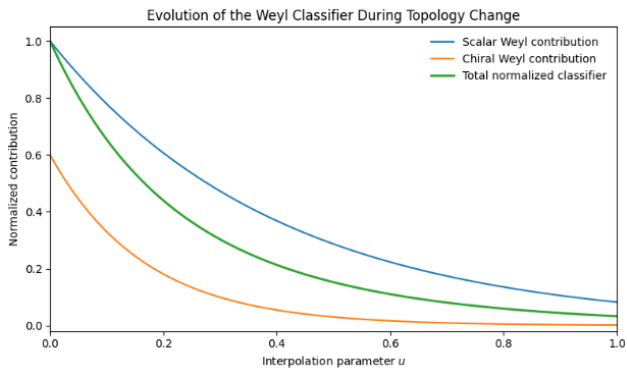

**Fig. 4.** Evolution of the normalized Weyl curvature diagnostic during topology change. Shown are the parity-even Weyl contribution (blue), the parity-odd contribution (orange), and the total diagnostic (green) as functions of the interpolation parameter $u$. All contributions decay smoothly as the geometry approaches the topologically trivial end state.

## VI. Conclusions

We have developed a geometric framework for modeling topology change in four-dimensional Lorentzian spacetimes using smooth interpolating spin cobordisms. By allowing a localized and temporary loss of global hyperbolicity—while preserving smoothness, Lorentz signature, and spin compatibility—the construction circumvents the classical Geroch–Tipler obstruction without introducing curvature singularities or closed timelike curves. The present results are intended primarily as geometric and conceptual diagnostics rather than phenomenological models

Within this framework, we identified conformal curvature as the dominant geometric response near wormhole throats and introduced a Chiral Weyl Wormhole Curvature Diagnostic as a covariant diagnostic of topology change. The Weyl curvature functional is built from the Weyl tensor and its parity-odd dual, rendering it sensitive to geometric chirality while remaining independent of coordinate choices. Spin admissibility is enforced via the vanishing of the second Stiefel–Whitney class, ensuring that only physically consistent cobordisms contribute.

The functional distinguishes mirror-related knotted embeddings, vanishes for amphichiral configurations, and decays smoothly as topology trivializes. While the constructions in this work focus on wormhole throats as a concrete and physically motivated setting, they should be regarded as representative examples rather than an exhaustive characterization of all possible Lorentzian topology-changing geometries. Replacing knot crossing number with a braid-generator–based complexity refines the normalization without altering the curvature structure of the diagnostic and extends its applicability to time-dependent, multithroat configurations.

The result is a curvature-level, chirality-sensitive, spin-filtered measure of the geometric cost of Lorentzian topology change. It is not a topological invariant and introduces no new dynamics; rather, it provides a principled geometric weighting suitable for regulating topology-changing configurations in semiclassical gravitational analyses. The framework developed here complements existing Euclidean and index-theoretic approaches by operating entirely within Lorentzian geometry and by resolving chirality at the level of curvature. A natural extension of the present framework is to examine how the parity-odd Weyl channel introduced here manifests in the fermionic sector, in particular through spectral asymmetry of Dirac operators on spin cobordisms; this direction will be developed separately and does not enter the results of the present work. In subsequent work, we show that this chirality-sensitive Weyl weight admits a natural algebraic interpretation in terms of oriented generator histories, providing a minimal combinatorial skeleton for Lorentzian topology change and we show that parity-odd Weyl curvature localized on spin-admissible cobordisms suggests a natural route for fermionic spectral asymmetry, providing a natural bridge from topology change to baryogenesis.

## Appendix A
## Cohomology Flow in Lorentzian Topology Change

Topology change in a Lorentzian spacetime can be characterized not only by geometric deformation but also by the evolution of global topological data encoded in homology and cohomology groups. In the present work, cohomology provides a complementary, kinematical perspective on the disappearance of wormhole-supported structure during a smooth spin cobordism [36, 37]. Each wormhole throat supports nontrivial cycles on spatial slices—typically contributing generators to $H_1$ and dual fluxes to $H_2$. As the cobordism proceeds, these cycles are smoothly contracted and capped off within the interpolating four-manifold. From the viewpoint of algebraic topology, this process corresponds to the annihilation of cohomology classes via the long exact sequence of the pair $(W, \Sigma)$. The "flow" of cohomology refers to this controlled disappearance of nontrivial classes across the cobordism. Importantly, this flow is global and discrete: unlike curvature, it does not decay continuously but instead terminates when the supporting cycles become null-homologous. The Weyl curvature diagnostic introduced in the main text may therefore be viewed as a geometric diagnostic whose vanishing coincides with the completion of this cohomological trivialization. In the context of quantum gravity and wormholes, this flow tracks the global topological features such as conserved fluxes, forms, or fields across a dynamical transition [38, 39, 40]. In the transition from a compact interacting multithroat wormhole configuration to a compact region of Minkowski spacetime $S^3$, the cohomology flow encodes the disappearance of nontrivial topological structure. Initially, each wormhole throat contributes nontrivial elements to $H^1$ or $H^2$ (e.g., representing fluxes threading the throats). As the cobordism proceeds, these cohomology classes are "killed off" by dual homology collapses — reflecting the cancellation or degeneration of the cycles supported by the wormhole geometry. Finally, in the Minkowski end-state, all cohomology beyond $H^3$ vanishes:

$$H^k\left(S^3\right) = \begin{cases} \mathbb{Z} & k = 0, 3 \\ 0 & \text{otherwise} \end{cases} \quad (A.1)$$

meaning all intermediate topological structure has flowed to triviality. The cohomology long sequence for the interpolating wormhole $\Sigma_1 = \Pi_i^N (S^1 \times S^2)_i$ topology to the compact $S^3$ Minkowski spacetime can be represented, with the inclusion induced morphisms i and δ as

$$H^0(S^2 \times S^1) = \mathbb{Z} \xrightarrow{i} H^0(W) \xrightarrow{\delta^*} H^1(W, S^2 \times S^1) \cdots H^1(S^2 \times S^1) = \mathbb{Z} \xrightarrow{i^*} H^1(W) \xrightarrow{\delta^*} H^2(W, S^2 \times S^1) \cdots$$
$$\cdots H^2(S^2 \times S^1) = \mathbb{Z} \xrightarrow{i^*} H(W) \xrightarrow{\delta^*} H^3(W, S^2 \times S^1) \cdots H^3(S^3) \quad .$$
(A.2)

This diagram tracks the time evolution of the cohomology groups $H^1(\Sigma(t))$ and $H^2(\Sigma(t))$ associated with spatial slices $\Sigma(t) \subset M$ of a wormhole-bearing spacetime $M$. These slices evolve over time $t$, during which wormhole throats form, interact, and disappear, resulting in topological transitions. Each horizontal row of the diagram represents a cohomology group at a given degree $k$, and each column represents the cohomology of a spatial slice $\Sigma t_i$ at a specific time $t_i$. Arrows indicate inclusion-induced morphisms between successive time slices.

Similarly, the long homotopic sequence for the pair $(W, \Sigma_1)$ is

$$\pi_2(\Sigma_1) = \mathbb{Z} \xrightarrow{i_*} \pi_2(W) \xrightarrow{j_*} \pi_2(W, \Sigma_1) \xrightarrow{\partial_*} \pi_1(\Sigma_1) = \mathbb{Z} \xrightarrow{i_*} \pi_1(W) \to \pi_1(W, \Sigma_1) \quad . \text{ (A.3)}$$

This diagram tracks the evolution of homotopy groups via the morphisms: $i_*$: inclusion maps, $j_*$: relative inclusion, and $\partial_*$: boundary homotopy maps. It shows the fundamental group decay from $\pi_1 = \mathbb{Z}$ to triviality and highlights where contractibility and obstructions occur. Viewed in this way, the cohomological evolution of the cobordism provides a global, discrete account of topological trivialization that complements the local curvature diagnostics used in the main text, without introducing additional dynamical assumptions. Note that the long exact sequence is used here diagnostically, to organize the response of curvature and fermionic invariants on spin-admissible cobordisms; the same cobordisms excluded by Lorentzian spin selection rules are assumed absent from the outset.

## Appendix B
## Braided Throat Dynamics, Generator Complexity, and Curvature Chirality

This appendix provides geometric motivation for the parity-odd Weyl curvature channel appearing in the Chiral Weyl Wormhole Curvature Diagnostic defined in Sec. V and makes explicit how braid data enter the normalization through a bounded complexity factor. The discussion is intentionally limited in scope: braid movies are used here as a minimal and geometrically transparent language for describing time-dependent wormhole throat dynamics on Lorentzian spin cobordisms. No claim is made that braid data furnish complete invariants of topology change.

### B.1 Braid movies as time-dependent throat representations
During a topology-changing spin cobordism $W$, wormhole throats evolve continuously in time. Tracking a fixed set of marked throat cross-sections through the interpolating region yields a collection of worldlines whose generic spacetime projections form a braid movie: a time-ordered sequence of local exchanges [41, 42]. In this representation, time runs vertically, while the ordering of throat strands defines the horizontal direction. The elementary events in a braid movie are encoded by generators $\sigma_i^{\pm 1}$ of the braid group $B_n$, representing localized over- and under-exchanges of neighboring throat segments. Because closed-braid representations are not unique—distinct braid words related by braid relations or Markov moves may represent the same link—braid movies are used here not as embedding invariants but as descriptive histories of local topology-changing events.

### B.2 Generator data and a bounded complexity factor
Let $B_n$ denote the braid group on $n$ strands with standard generators $\{\sigma_1, \dots, \sigma_{n-1}\}$. A braid word
$$\beta = \sigma_{i_1}^{\epsilon_1} \sigma_{i_2}^{\epsilon_2} \cdots \sigma_{i_m}^{\epsilon_m}, \epsilon_k \in \{+1, -1\}, \tag{B.1}$$

has word length $\ell(\beta) = m$. Let $\hat{\beta}$ denote its closure, representing the instantaneous throat configuration on a spatial slice. To incorporate braid information into the Weyl curvature diagnostic without introducing algebraic overreach, we define a bounded braid complexity factor that depends only on readily computable braid data [43, 44].
Definition B.1 (Braid complexity factor). For a throat configuration $L$ represented as the closure $L = \hat{\beta}$ of some braid $\beta \in B_n$, define
$$\mathcal{K}_{\mathrm{br}}(L) \equiv 1 + \alpha(n-1) + \gamma \ell(\beta), \tag{B.2}$$

where $\alpha, \gamma > 0$ are fixed dimensionless constants. The braid $\beta$ may be chosen as any representative produced by the chosen throat-tracking or braid-movie extraction procedure. The role of $\mathcal{K}_{\mathrm{br}}$ is purely normalizing it weights the Weyl curvature diagnostic by the combinatorial richness of the braid representation while remaining finite and insensitive to geometric rescaling. No minimization over Markov-equivalent representatives is assumed.

### B.3 Generator-level chirality

The braid word $\beta$ determines a signed sequence of generator events. This motivates a corresponding measure of generator chirality. Definition B.2 (Signed generator sum): for a braid word $\beta$ as above, define

$$\chi(\beta) \equiv \sum_{k=1}^{\ell(\beta)} \epsilon_k. \tag{B.3}$$

The quantity $\chi(\beta)$ changes sign under braid inversion and vanishes for words in which generator events occur in orientation-reversed pairs. In particular, braid representations of amphichiral embeddings—such as the figure-eight knot—necessarily admit a pairing of generators with opposite signs, forcing $\chi(\beta) = 0$; this algebraic cancellation mirrors the vanishing of the parity-odd Weyl contribution observed for amphichiral throats in Sec. V.

### B.4 Relation to the parity-odd Weyl channel

The parity-odd contribution to the Weyl curvature diagnostic introduced in Sec. V is governed by the dual Weyl contraction $C_{abcd}{}^* C^{abcd}$, which changes sign under orientation reversal and vanishes for mirror-symmetric geometries. In the braid-movie picture, each generator event $\sigma_i^{\pm 1}$ corresponds to a localized, oriented deformation of the geometry. This motivates the schematic interpretation

$$\int_W C_{abcd}{}^* C^{abcd} \sqrt{-g}\, d^4x \ \sim \ \sum_k \epsilon_k \, \mathcal{A}_k, \tag{B.4}$$

where $\mathcal{A}_k$ denotes the spacetime-weighted contribution of the $k$-th local exchange event to the parity-odd curvature response. Equation (B.3) is not a claim of equality or quantization. Rather, it provides a geometric bookkeeping interpretation consistent with (i) sign reversal under orientation reversal and (ii) exact cancellation for amphichiral braid histories.

### B.5 Insertion into the Weyl curvature diagnostic normalization

With the braid complexity factor defined above, the normalization $N$ appearing in Eq. (13) of Sec. V may be written in the form

$$N = (1 + L/L_0)\, \mathcal{K}_{\mathrm{br}}(L), \tag{B.5}$$

where $L$ is the proper arclength of the wormhole throat and $L_0$ is a fixed reference scale. This makes explicit how braid index and generator count enter the Weyl curvature diagnostic only through normalization, without altering the curvature content or chirality sensitivity of the Weyl functional itself.

### B.6 Outlook: toward a group-theoretic formulation of topology change

The braid-movie picture adopted here naturally suggests a broader group-theoretic formulation of topology change. In the present work, braid generators serve as a concrete and geometrically transparent example of elementary topology-changing events associated with wormhole throat dynamics. More generally, one may regard evolving throat networks as histories in a discrete group or category of local topology-change generators, with braid groups providing the simplest nontrivial realization. A systematic treatment of topology change in terms of group generators, relations, and representations — including extensions beyond braid groups and their connection to curvature and spin structure — will be developed separately. The present paper uses braid movies only as a minimal and illustrative realization of this more general perspective.

<div align="center">

## Appendix C
## Energy and Curvature during the topology Change

</div>

This appendix provides supporting calculations for the interpolating geometry used in the main text, confirming smoothness and boundedness of curvature invariants and illustrating how standard energy-condition diagnostics evolve during the topology-changing interpolation. The results here are intended as consistency checks and illustrative correlations for the example family of metrics, rather than as general statements about matter requirements for topology change. We evaluate representative curvature scalars and stress-energy components along the cobordism parameter $u$, with the goal of (i) confirming smoothness and boundedness of curvature invariants throughout the interpolation, and (ii) illustrating how standard energy-condition diagnostics evolve in the same regime where the curvature-based functional of Sec. V is nontrivial [45, 46, 47]. We write the energy density and principal pressures as functions of the radial coordinate $r$ and the interpolation parameter $u$, with $u = 0$ corresponding to the Morris–Thorne limit and $u = 1$ corresponding to the asymptotic vacuum end-state of the interpolation.

$$\begin{aligned}
&NEC: R_{\mu\nu}k^{\mu}k^{\nu} \geq 0 \rightarrow T_{\mu\nu}k^{\mu}k^{\nu} \geq 0 \rightarrow \rho - p_r \geq 0, \quad g_{\mu\nu}k^{\mu}k^{\nu} = 0 \\
&WEC: R_{\mu\nu}u^{\mu}u^{\nu} \geq 0 \rightarrow T_{\mu\nu}u^{\mu}u^{\nu} \geq 0 \rightarrow \rho + p_r \geq 0 \quad \rho + p_\theta \geq 0 \quad \text{and} \quad \rho \geq 0, \quad g_{\mu\nu}u^{\mu}u^{\nu} < 0 \\
&SEC: \left(R_{\mu\nu} - \frac{1}{n-2}Rg_{\mu\nu}\right)u^{\mu}u^{\nu} \geq 0 \rightarrow \left(T_{\mu\nu} - \frac{1}{n-2}Tg_{\mu\nu}\right)u^{\mu}u^{\nu} \geq 0, \quad \rho + p_r + 2p_\theta \geq 0 \\
&DEC: \left(R_{\mu\nu} - \frac{1}{n-2}Rg_{\mu\nu}\right)u^{\mu}u^{\nu} \geq 0, \quad \left(R_{\mu\nu} - \frac{1}{n-2}Rg_{\mu\nu}\right)u_k^{\mu} \geq 0, \quad T_{\mu\nu}u^{\mu}u^{\nu} \geq 0, \quad T_{\mu\nu}u_k^{\mu} \geq 0, \quad \rho \geq 0, \rho \geq |p_r|, \rho \geq |p_\theta|
\end{aligned} \quad (C.1)$$

For our wormhole metric the energy momentum tensor determines the density and pressure terms which can be expressed in terms of the cobordism parameter u as

$$\rho = \frac{1}{8\pi}\left[\frac{(1-u)b'(r)}{r^2} - \Lambda r^2\right] \quad (C.2)$$

where u=0 gives the standard Morris Thorne wormhole values and u=1 gives the Minkowski spacetime with a cosmological constant. The remaining pressure components and combinations entering the NEC and WEC follow analogously and are evaluated numerically for the examples shown below. In order to not violate the WEC in the wormhole geometry the cosmological constant acts as a regulator and is constrained by

$$\rho: \Lambda \geq \frac{r_0^2}{r^4}, \quad p_r: \Lambda \geq \frac{r_0^2}{r^4}, \quad p_\theta: \Lambda \geq \frac{r_0^2}{2r^4}(r+1) \quad . \quad (C.3)$$

Here the cosmological constant enters only as a background curvature scale and is constant during the interpolation. Its purpose is to separate infrared curvature effects from the localized Weyl-dominated response of the wormhole throat, ensuring that violations of energy conditions near the throat are not conflated with asymptotic behavior. Setting $\Lambda \rightarrow 0$ recovers the corresponding asymptotically flat limit without altering the qualitative features of the topology-changing transition. Indicating that a large enough cosmological constant could in principle make $\rho$, $\rho + p_r$ and $\rho+p_\theta$ non-negative avoiding the WEC. For large r the cosmological constant term can easily dominate and satisfy the WEC, however, near the throat the cosmological constant term becomes smaller, and it is not always possible to satisfy the WEC without considerable fine tuning. In general, an envelope function such as

$$\rho(r,u) = \Lambda - \frac{r_0^2(u)}{r^4} - \frac{A(u)}{r^2\sigma^2}(r-r_0(u))\exp\left(-\frac{(r-r_0(u))^2}{2\sigma^2}\right) \quad (C.4)$$

is introduced where the minimum throat radius is a function of u. The energy violations for the NEC and WEC cases are plotted in Fig.(C.1) for the radial and tangential pressure terms of a wormhole with a cosmological constant. They all vanish in Minkowski spacetime where u=1. Here we plot the energy density, radial and tangential pressures, exotic matter density, and curvature scalars during the dynamical topology change.

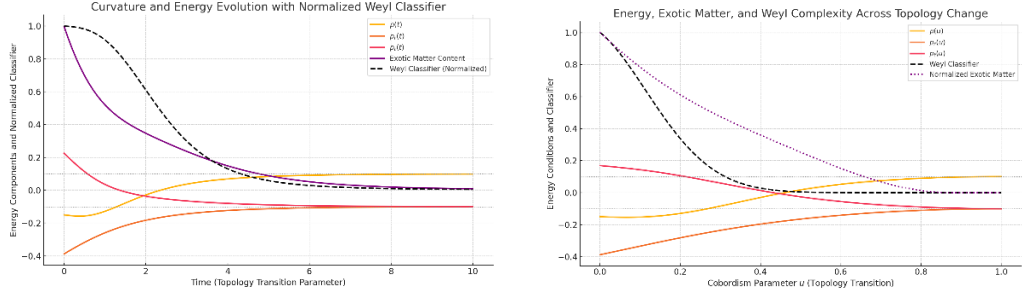

**Fig. C.1.** *Curvature and Energy Density* (a) Curvature and Weyl for several knots, (b) topology change. This illustrates the evolution of energy density, radial and tangential pressures, exotic matter content, and the Weyl curvature functional during a smooth topology-changing transition from a trefoil-knotted wormhole to a topologically trivial $S^3$ geometry.

This results in the wormhole's geometry being modified by a Gaussian envelope modeling the localized curvature of a trefoil knot, which enhances both conformal curvature and pressure anisotropy near the throat. Early in the transition, negative energy density and large radial tension signify NEC violation and the presence of exotic matter necessary to support the knotted structure. As the cobordism progresses, the knot collapses, the total exotic matter content decays to zero, with stress-energy components approaching the de Sitter vacuum form $(\rho, p_r, p_\theta) = (\Lambda, -\Lambda, -\Lambda)$ in the asymptotic $u \to 1$ limit. This behavior is consistent with the disappearance of localized geometric obstructions as the cobordism approaches the trivial end-state. The Weyl Complexity Curvature Diagnostic, which captures the combined effect of curvature concentration, knot complexity, and spin structure, rises during the knotted phase and decays as the wormhole dissolves. In summary, the examples presented here confirm that the interpolating Lorentzian spin cobordisms remain smooth and free of curvature singularities throughout the topology-changing transition. Energy-condition violations and exotic matter contributions are confined to the localized throat region and decay in tandem with the Weyl curvature diagnostic as the geometry relaxes to the topologically trivial end state.